\documentclass[11pt]{article}

\usepackage[margin=1in]{geometry}
\usepackage[utf8]{inputenc}
\usepackage[T1]{fontenc}
\usepackage{times}
\usepackage{amsmath}
\usepackage[compress]{cite}
\usepackage{booktabs}
\usepackage{array}
\usepackage{longtable}
\usepackage{ragged2e}
\usepackage[hidelinks]{hyperref}
\usepackage{titlesec}
\usepackage{enumitem}
\usepackage{parskip}
\usepackage{microtype}

\titleformat{\section}{\normalfont\Large\bfseries}{\thesection}{1em}{}
\titleformat{\subsection}{\normalfont\large\bfseries}{\thesubsection}{1em}{}

\title{\textbf{A Protocol for Evaluating the Accessibility of AI-Generated Educational Materials: Prompt Configuration, WCAG-Derived Criteria, and Content Overload}}
\author{Hector R. Amado-Salvatierra \\ \small Universidad Galileo \\ \small \href{mailto:hr_amado@galileo.edu}{hr\_amado@galileo.edu} \\ \small ORCID: \href{https://orcid.org/0000-0002-3028-234X}{0000-0002-3028-234X}}
\date{}

\begin{document}

\maketitle

\begin{abstract}
\noindent
Generative AI tools increasingly produce educational materials: documents, slide presentations, images and infographics, narrated audio, and video. Little is known about whether this content meets established accessibility requirements. This paper presents a protocol for evaluating the accessibility of AI-generated educational materials against the Web Content Accessibility Guidelines (WCAG), across five content types and multiple generative AI tools. The protocol compares three conditions applied to the same tool: a generic, everyday instruction with no accessibility language; a single prompt explicitly configured with operationalized WCAG criteria; and a persistent, reusable accessibility profile (``skill'') loaded once into the authoring tool rather than re-specified for each request. Evaluation combines an operationalized WCAG rubric per content type with heuristic validation by a panel of accessibility experts, addressing a documented limitation of automated accessibility scanners. Prior evidence shows that generative AI tools reproduce inaccessible practices by default, and that explicit configuration measurably improves compliance. This paper extends that discussion to a dimension conventional WCAG checklists do not capture: the visual and informational overload frequently observed in AI-synthesized content, illustrated through the case of NotebookLM-style summarization tools. It also proposes a persistent configuration approach as a candidate solution to the inconsistency of one-off prompting. The main contribution is methodological: a reproducible, content-type-specific protocol and an open evaluation instrument that other researchers and practitioners can apply to their own generative AI workflows, together with illustrative use cases documenting the specific barriers that non-configured AI-generated content creates for people with disabilities. A first exploratory application of the protocol is also reported, using the tools available at the time of writing: a rubric-based compliance score rose from a pooled mean of 24.2\% under the generic condition to 96.7\% under the WCAG-configured condition. The full protocol specifies three experimental conditions, a generic prompt with no accessibility language, a single prompt explicitly configured with WCAG criteria, and a persistent accessibility profile loaded once into the tool, together with expert-panel validation. This first exploratory application was limited to the first two of these conditions, the generic and the WCAG-configured prompt, using one general-purpose model, one artifact per condition and content type, and single-evaluator rubric scoring. The protocol itself, together with its open evaluation instrument, is proposed as a reusable resource for the community to apply and extend at full benchmark scale. Beyond the protocol itself, the broader aim of this work is to raise awareness of the accessibility obstacles generative AI tools can introduce, and to encourage content creators to take accessibility into account whenever they use these tools to produce educational materials.

\vspace{0.5em}
\end{abstract}

\section{Introduction}

Generative artificial intelligence tools have rapidly become part of how instructors, instructional designers, and content developers produce educational materials, from written explanations and slide decks to narrated audio and short instructional videos \cite{bib26,bib27}. These tools promise to reduce the time and specialized skill previously required to produce this kind of content, and prior work on personalized feedback pipelines and AI-supported content creation in massive open online courses has already documented measurable gains when generative components are properly orchestrated rather than used as a single unstructured request \cite{bib25,bib27}. However, this shift raises a question that has received comparatively little systematic attention: whether the materials these tools produce are accessible to learners with disabilities.

Accessibility of digital content has an established normative framework. The Web Content Accessibility Guidelines (WCAG), now in version 2.2 and adopted as the international standard ISO/IEC 40500:2025, organize accessibility requirements around four principles: content must be Perceivable, Operable, Understandable, and Robust \cite{bib1}. These requirements are not new, and a research tradition spanning more than a decade has applied them to e-learning specifically: embedding accessibility standards directly into the course-design workflow \cite{bib20}, evaluating the accessibility of major MOOC platforms \cite{bib21}, training technical staff in virtual learning environments to apply accessibility criteria as part of their daily work \cite{bib22}, and designing fully accessible online courses as demonstration cases \cite{bib23,bib24}. What generative AI changes is not the normative framework, but the nature of who, or what, now produces the content that must comply with it.

However, a growing body of evidence indicates that generative AI tools do not meet these requirements by default. Studies evaluating AI-generated web code have found accessibility violations in the large majority of outputs produced without explicit instruction \cite{bib2,bib4,bib30}, and this pattern extends to documents, images, and other content types. The reason is structurally similar across all of them: models trained on a web where an estimated 94.8\% of home pages already fail at least one WCAG success criterion \cite{bib33} tend to reproduce, rather than correct, the inaccessible practices present in their training data \cite{bib28}. At the same time, a smaller but consistent line of work shows that this is not an inherent limitation of the technology: when generative AI tools are given explicit, operationalized accessibility instructions, through carefully engineered prompts \cite{bib6,bib29}, structured remediation pipelines \cite{bib7,bib10}, or context-aware generation conditioned on the surrounding material \cite{bib8}, measured compliance improves substantially compared to unconfigured use of the same tool. This paper's central premise follows directly from this contrast: AI-generated educational content is not accessible by default, but it can become substantially more accessible when the generating tool is deliberately configured, and the practical question this raises is how that configuration should be delivered and sustained.

A further complication, less discussed in the accessibility literature specifically, is that AI tools designed to synthesize source material into compact visual formats, as illustrated by the case of NotebookLM-style tools, can introduce a distinct barrier that a binary WCAG checklist does not capture: information and visual overload. Prior analysis of this class of tool has shown that its underlying compression mechanism differs fundamentally from human reading and analysis, at times leaving out important content from the source or adding material that was not actually present in it \cite{bib15}; the visual outputs of such tools compound this with dense, unstructured infographics that are difficult to process even for sighted users, independent of whether a text alternative is present.

Given this context, this paper proposes and applies a review protocol to answer five research questions:

\begin{itemize}[leftmargin=1.5em]
\item[] \textbf{RQ1:} To what extent does AI-generated educational content comply with operationalized WCAG criteria under default, unconfigured use?
\item[] \textbf{RQ2:} Does a single prompt explicitly configured with accessibility criteria measurably improve compliance, compared to a generic prompt, on the same tool?
\item[] \textbf{RQ3:} Which content type, documents, slide presentations, images/infographics, audio, or video, shows the largest improvement under configuration, and which shows the most persistent barriers?
\item[] \textbf{RQ4:} To what extent do AI synthesis tools, illustrated by the NotebookLM case, introduce visual/informational overload as a distinct accessibility barrier not captured by conventional WCAG criteria?
\item[] \textbf{RQ5:} Does a persistent, reusable accessibility configuration profile, analogous to a saved ``skill'' or system-level instruction set loaded once into the authoring tool, produce more consistent WCAG compliance across multiple, varied content-generation requests than a single ad-hoc configured prompt repeated each time?
\end{itemize}

The remainder of this paper is organized as follows. Section~2 details the WCAG normative framework and its four principles, discusses the potential and the accessibility risk of the speed at which generative AI now produces instructional content, and reviews related work on the accessibility of AI-generated code, images, text, and documents, on the role of prompt configuration, and on content overload in AI synthesis tools. Section~3 presents the review protocol, including the content types, tools, three experimental conditions, the operationalized WCAG evaluation instrument, and the expert heuristic validation procedure. Section~4 reports illustrative results for each content type, together with the compliance comparison from a first exploratory application of the protocol and the design for the full benchmark it proposes. Section~5 discusses these results in light of the literature reviewed in Section~2. Section~6 states the limitations of this stage of the work, and Section~7 outlines future work directly addressing them. Finally, Section~8 presents the conclusions.

\section{Related Work}

\subsection{Understanding WCAG: the POUR principles and what non-compliance means in practice}

The Web Content Accessibility Guidelines have gone through four major revisions since their first publication in 1999: WCAG 1.0, then WCAG 2.0 in 2008, WCAG 2.1 in 2018, and WCAG 2.2 in 2023. Each round adds success criteria without deprecating the previous ones, so a document or interface compliant with 2.2 is, by construction, also compliant with the earlier versions it extends \cite{bib1}. The 2.2 revision was subsequently adopted as ISO/IEC 40500:2025, which confirms its normative weight well beyond a web-specific recommendation and gives institutions and governments a standard they can cite in procurement and accessibility policy, not only in web design guidelines \cite{bib1}. Compliance is assessed at three levels, A, AA, and AAA, in increasing order of strictness. Most institutional policies, including the target this protocol assumes (Section~3.3), require Level AA, which is also the level referenced throughout the empirical literature reviewed below.

WCAG organizes its success criteria around four principles, commonly abbreviated POUR, and each principle addresses a different way that content can become unusable rather than merely inconvenient. Table~\ref{tab:pour} summarizes the four principles, the guideline areas each one covers, and an illustrative barrier that a learner encounters when that principle is not met, these examples are not drawn from a specific empirical study, but from the documented failure modes WCAG itself is designed to prevent, and they recur, in one form or another, throughout the AI-generated content literature reviewed in Sections~2.4--2.7.

\begin{table}[htbp]
\centering
\small
\caption{WCAG's four principles (POUR), representative guideline areas, and illustrative accessibility barriers.}
\label{tab:pour}
\begin{tabular}{p{2.0cm}p{5.3cm}p{6.3cm}}
\toprule
\textbf{Principle} & \textbf{Representative guideline areas} & \textbf{Illustrative barrier if unmet} \\
\midrule
Perceivable & Text alternatives for non-text content; captions and other alternatives for audio/video; content adaptable to different presentations without losing structure; sufficient contrast and distinguishability of text from background & A blind learner using a screen reader hears nothing where a chart should be described, because the chart has no text alternative; a low-vision learner cannot read body text set against a background with insufficient contrast \\
\addlinespace
Operable & All functionality available from a keyboard; enough time to read and use content; no content that causes seizures or physical reactions; navigable structure with clear headings, labels, and focus order & A learner with a motor disability who cannot use a mouse is unable to open an interactive quiz embedded in a slide deck, because none of its controls are reachable by keyboard \\
\addlinespace
Understandable & Readable and predictable text; consistent navigation and identification; input assistance and error prevention in forms & A learner with a cognitive or learning disability is unable to follow a densely worded, jargon-heavy explanation that could have been expressed in plain language, or is confused by navigation that behaves inconsistently across pages \\
\addlinespace
Robust & Compatible with current and future assistive technologies, including correct semantic markup that such technologies can parse & A deaf or hard-of-hearing learner relying on browser-based live captioning finds that a video player is not exposing its media controls in a way the captioning extension can read, silently breaking the accessibility feature \\
\bottomrule
\end{tabular}
\end{table}

In practice, non-compliance is rarely a single missing feature; it is usually a chain in which one unmet guideline (e.g., no text alternative) produces a downstream barrier (e.g., a screen reader has nothing to announce) that a learner experiences as content simply not working, regardless of how well-designed the underlying instructional material might otherwise be. This chain is what the four illustrative use cases in Section~4 document concretely for AI-generated content specifically.

\subsection{A decade of accessible e-learning research}

Universal Design for Learning (UDL) complements this technical framework with a pedagogical one, arguing that flexible means of representation, expression, and engagement benefit the full range of learners, not only those with documented disabilities \cite{bib19}. This paper builds on a research program that applied both frameworks to e-learning before the emergence of generative AI. Early work embedded WCAG-based accessibility standards directly into the course-design process of virtual learning environments, showing that compliance improves when standards are built into the authoring workflow rather than checked afterward \cite{bib20}. Subsequent work evaluated the accessibility of major MOOC platforms against automated accessibility checkers \cite{bib21} and trained technical staff in virtual learning environments to apply accessibility criteria as part of their routine work \cite{bib22}. Project-level work under the ESVI-AL initiative proposed a broader institutional approach to accessible and inclusive virtual education \cite{bib23}, and a fully accessible open online course was designed and evaluated as a demonstration case of compliant e-learning content in practice \cite{bib24}. What changes with generative AI is not this framework or its workflow-embedding logic, but the nature of who, or what, produces the content that must comply with it, and, as Section~2.3 argues, the speed at which that content is now produced.

\subsection{Generative AI in instructional design and content production: potential and the speed-accessibility tension}

Generative AI tools are changing the economics of instructional design in a way earlier authoring tools did not. Where producing a slide deck, a narrated module, or a short explainer video once required either specialized multimedia skills or a commissioning process involving instructional designers and media producers, a single natural-language prompt can now generate a full first draft of any of these formats in minutes. Work combining AI-based tools with human pedagogical decision-making during MOOC design found that this can measurably shorten the design cycle without sacrificing quality, provided instructors retain oversight over pedagogical choices rather than accepting AI output unmodified \cite{bib25}. Personalized feedback pipelines built by explicitly orchestrating language-model components, rather than issuing a single unstructured prompt, produced more consistent, higher-quality output for learners at a scale that would be impractical for a human tutor to sustain manually \cite{bib27}. At the level of content creation specifically, the deliberate design of prompts has been identified as a core competency educators need to obtain usable results when generating instructional materials with large language models, precisely because the same underlying model can produce dramatically different output quality depending on how the request is framed \cite{bib26}.

However, this same property, that a well-crafted prompt can produce a polished-looking deliverable almost instantly, is what creates the tension this paper investigates. Traditional content production, however slow, has historically passed through stages (storyboarding, editing, a design review) where an accessibility check could plausibly be inserted, even if it often was not. Generative AI collapses many of those stages into a single generation step, which removes natural checkpoints unless accessibility is deliberately built back into the prompt or the workflow. Table~\ref{tab:production} contrasts these two production modes across dimensions directly relevant to accessibility risk.

\begin{table}[htbp]
\centering
\small
\caption{Traditional versus AI-assisted educational content production: dimensions relevant to accessibility risk.}
\label{tab:production}
\begin{tabular}{p{2.6cm}p{3.4cm}p{3.6cm}p{4.0cm}}
\toprule
\textbf{Dimension} & \textbf{Traditional/manual production} & \textbf{AI-assisted production (unconfigured)} & \textbf{Accessibility implication} \\
\midrule
Time to produce one module or asset & Hours to days, often involving specialized staff & Minutes, from a single prompt & Faster iteration means more assets are produced per unit of instructor time, multiplying the consequence of any recurring accessibility defect \\
\addlinespace
Required authoring skill & Instructional design and, often, dedicated multimedia/accessibility expertise & A natural-language request; no specialized skill required by default & Removes a historical bottleneck that also functioned, informally, as a quality/accessibility checkpoint \\
\addlinespace
Natural review stage & Storyboard, draft, and design-review stages where accessibility could be checked & A single generation step, unless the workflow explicitly reintroduces a review stage & Fewer natural opportunities for manual accessibility remediation unless deliberately re-inserted \\
\addlinespace
Reuse/adaptation of prior material & Templates and style guides accumulate accessibility fixes over time within an institution & Each generation can be a fresh, stateless request unless a persistent profile (Condition C, Section~3.2) is used & Institutional accessibility knowledge does not automatically carry over between requests without an explicit mechanism \\
\addlinespace
Volume of content producible per instructor & Limited by staff time and budget & Practically unbounded, limited mainly by review capacity & Scales the total amount of potentially inaccessible content much faster than institutional accessibility review capacity typically scales \\
\bottomrule
\end{tabular}
\end{table}

Given this contrast, the central risk this paper addresses is not that generative AI tools are incapable of producing accessible educational content (Section~2.5 reviews evidence that they can, when configured). The risk is that the same feature that makes them valuable for instructional design, their speed and low barrier to use, allows inaccessible defaults to propagate at a scale and pace that outstrips the review capacity most institutions have historically relied on. This motivates treating configuration (RQ2), and more specifically persistent configuration (RQ5), as central research questions rather than secondary ones.

\subsection{AI-generated content does not comply with accessibility requirements by default}

Empirical studies converge on a central finding: unless explicitly instructed, content produced by generative AI tools does not meet accessibility requirements by default. Aljedaani et al.\ \cite{bib2} evaluated 88 websites generated by ChatGPT and found accessibility issues in 74 of them (84\%). Mapped onto the POUR principles in Table~\ref{tab:pour}, this proportion was dominated by Perceivable and Robust violations (missing text alternatives and markup that assistive technologies could not reliably parse), rather than an even spread across all four principles. Ahmed et al.\ \cite{bib3} showed that GPT-4o, generating a web page without accessibility instructions, omits elements such as the ``main'' landmark or a skip-to-content link. It corrects these when explicitly and iteratively prompted, and supplying screenshots alongside text prompts further improved the model's ability to reason about contrast and surrounding layout. Abu Doush and Kassem extended this comparison across four models and eleven web components \cite{bib4}, and in a related metric-driven evaluation \cite{bib30}. Performance was uneven, with no model achieving consistent compliance unassisted regardless of which of the four models was tested. Palmer and Oswal reached a converging conclusion analyzing the full workflow of AI-assisted website construction: the resulting product presents persistent barriers unless the process incorporates manual review \cite{bib5}. Alshaigy and Grande situate this within a broader pattern, arguing that accessibility considerations continue to be ``forgotten again'' in the design of new generative AI tools despite well-documented evidence of their importance \cite{bib28}. This framing anticipates the argument in Section~2.3 that production speed, not malicious intent or technical incapacity, is what allows this gap to persist.

This pattern is not limited to code. Alt-text generation for complex technical or educational images, diagrams, charts, equations, remains prone to missing or low-quality descriptions \cite{bib11}. Cardia et al.\ report that even when LLM-generated alt-text for STEM images passes a superficial similarity check against human-written descriptions, it more often omits the specific numeric or structural detail (an axis label, a data point, a step in a diagram) that makes the description usable for someone who cannot see the image at all \cite{bib11}. PDF documents present a parallel case: correct tagging, reading order, and accessible descriptions of tables and mathematical notation require more than unguided automated processing. Kumar, Padath, and Wang built a benchmark dataset of expert-annotated scholarly PDFs specifically to test whether large language models can support this evaluation, scoring documents across seven distinct criteria: alt-text quality, logical reading order, semantic tagging, table structure, functional hyperlinks, color contrast, and font readability. Automated tools disagreed with expert human annotation most often on reading order and table structure, the two criteria that require understanding the document's logical layout rather than just its visible text \cite{bib32}.

\subsection{The role of configuration: when AI does support accessibility}

The complementary finding is that these tools can support concrete accessibility tasks when properly configured, and the size of the improvement reported across this literature is large enough to treat configuration as the central independent variable of this paper's protocol (Condition A versus Condition B, Section~3.2), not a minor refinement. Vera-Amaro and Rojano-C\'aceres showed that specific prompting strategies and template-guided remediation measurably improve accessibility compliance compared to generic instructions on the same model \cite{bib6}, and a dedicated system, ACCESS, automates web accessibility correction through structured, iterative prompt engineering against the live document object model, inspecting the rendered page, identifying specific violations, and issuing targeted correction prompts in a loop, rather than a single static instruction \cite{bib29}. A step-by-step, guided remediation workflow for PDF tagging nearly doubled tagging accuracy for both experienced and novice users compared to unguided use of a standard tool \cite{bib10}, reinforcing that structured configuration, not raw model capability, drives accessible output. Dedicated pipelines extend this logic to alt-text generation: AltGen reduced accessibility errors by 97.5\% relative to unguided generation through a structured extraction-and-context pipeline \cite{bib7}, and a related system generates alt-text for images in educational documents by conditioning generation on surrounding context rather than the image alone, addressing exactly the ``purpose blindness'' limitation Cardia et al.\ documented in Section~2.4 \cite{bib8}. A 2026 model-based approach goes a step further, generating accessible interfaces directly from an explicit accessibility model consulted by the LLM during generation, rather than from prompt text alone \cite{bib31}, the model is consulted at each generation step rather than supplied once as instruction text, which is conceptually close to what this paper's Condition C tests with a persistent profile, though implemented at the level of the generation architecture rather than the prompt layer, and is a framing directly relevant to RQ5's persistent-profile approach.

Taken together, this evidence shows that the improvement from configuration is not confined to one content type or one correction mechanism. A single well-engineered prompt \cite{bib6}, an iterative DOM-level correction loop \cite{bib29}, a step-by-step guided workflow \cite{bib10}, a context-conditioned generation pipeline \cite{bib7,bib8}, and an explicit accessibility model consulted during generation \cite{bib31} all report substantial gains over unconfigured use, despite targeting different content types and using different mechanisms. This convergence across otherwise unrelated studies justifies treating ``configuration helps'' as an established premise (stated in Section~1) rather than as RQ2's open question. What remains open, unanswered by any single one of these studies, is whether the gain holds consistently across five different content types evaluated under one shared protocol (RQ2--RQ3), and whether it can be sustained across repeated use without re-specifying the same instructions each time (RQ5).

\subsection{Readability, text simplification, and captioning}

A related line of work addresses the textual and audiovisual accessibility of generated content rather than WCAG multimedia compliance directly. Dedicated readability benchmarks support automated readability assessment, including work specifically targeting Spanish-language text, which matters for any future extension of this protocol beyond English-only materials \cite{bib12}. Subsequent work has explored generating Easy-to-Read content, a simplified register defined by disability advocacy standards for readers with cognitive or intellectual disabilities \cite{bib13}, and adapting text to plain language through automatic post-editing cycles that iteratively simplify a draft rather than generating simplified text directly on the first attempt \cite{bib14}. Captioning research shows a parallel pattern for audiovisual content: automatically generated captions can measurably support learning comprehension and reduce cognitive load, but the benefit depends on caption quality and language pairing, not merely on captions being present, in other words, a caption track is necessary but not sufficient, echoing this paper's broader argument that presence of a feature and quality of that feature are distinct questions \cite{bib18}. However, none of this work integrates the multimedia WCAG criteria, semantic structure, alt-text, document tagging, central to the tradition summarized in Section~2.1, treating readability and WCAG compliance as separate research problems even though both affect the same learners.

\subsection{Content overload and visual saturation: the NotebookLM case}
\label{sec:overload-related}

An aspect largely absent from WCAG-focused literature is the visual and informational overload produced by AI synthesis tools, even where no strict technical criterion is violated. Analysis of NotebookLM's use in educational contexts argues that its statistical compression mechanism differs fundamentally from human reading and analysis, at times leaving out arguments from the source material or introducing content that was not actually present in it \cite{bib15}. This is compounded by documented technical accessibility issues, incomplete screen-reader and keyboard support, visual elements that clash under screen magnification, and inconsistent alternative descriptions for source images, and by a more informally documented pattern in which automatically generated infographics overload the visual output with dense, unclear hierarchy, hindering comprehension independent of whether alt-text is present. This is precisely the failure mode Table~\ref{tab:pour} does not capture in its Perceivable row: an infographic can have a complete, accurate text alternative and still be perceivable in the narrow WCAG sense while remaining unusable for a learner who cannot process a visually dense image quickly, which is why RQ4 treats overload as its own dimension rather than folding it into the existing POUR framework.

\subsection{Frameworks for evaluating generative AI in educational materials}

A separate line of work proposes rubrics for the educational value of generative AI output. The Edu-GenAI Rubric organizes evaluation across five domains, Instrumental, Technical, Hedonic, Use, and Beneficial value, covering dimensions from accuracy and productivity to data privacy and algorithmic fairness, intended to support institutional adoption decisions rather than accessibility auditing specifically \cite{bib16}. A separate comparison of multiple LLMs used as automated judges against instructional-materials rubrics found that model-based judging can approximate human expert ratings on general instructional quality, but the study did not test whether the same automated-judge approach could reliably assess accessibility-specific criteria, which require checking underlying markup and structure rather than the surface quality of the content a judge model can read directly \cite{bib17}. None of these rubrics integrates WCAG criteria as a central evaluation axis, nor treats content/visual overload as a distinct dimension, which is the specific combination this paper's evaluation instrument (Table~\ref{tab:criteria}, Section~3.3) is designed to provide.

\subsection{What tool providers claim about accessibility}
\label{sec:provider-claims}

The premise that AI-generated content is not accessible by default (RQ1) is easy to state in the abstract, but it is worth checking directly against what the providers of the tools named in Section~3.2 themselves claim, since the picture is not uniform. Two claims need to be kept apart. A provider can document that its own chat or editor \textit{interface} is usable by people with disabilities, typically via a Voluntary Product Accessibility Template (VPAT) or Accessibility Conformance Report (ACR); this says nothing about whether \textit{content generated through} that interface meets WCAG. Alternatively, a provider can market explicit features intended to make the \textit{output} more accessible, which is the claim most directly relevant to RQ1--RQ2.

Table~\ref{tab:toolaccess} summarizes what was found, as of July 2026, for every tool named in the frozen-versions record (Section~3.2 and the companion versioning file). The pattern is uneven along exactly the interface-versus-output distinction above. The three general-purpose assistants used across content types, ChatGPT, Claude, and Gemini, either lack a public accessibility report altogether (ChatGPT) or publish one for their own interface only (Claude's Enterprise iOS ACR; Gemini's web VPAT), with no equivalent claim that a document, slide deck, or image a user asks them to generate will itself be WCAG-conformant. NotebookLM, in particular, has documented interface-level accessibility gaps of its own, reported independently by university digital-accessibility offices, at the same time that it is the case study this paper uses for content overload (Section~\ref{sec:overload-related}). By contrast, three of the specialized authoring tools, Genially, Canva, and HeyGen, explicitly market built-in features meant to help a creator produce accessible \textit{output} (a design accessibility checker, captioning/audio-description templates). Even here, the providers qualify the claim: Genially frames accessibility of published content as a shared responsibility between the platform and the creator, and Canva's own accessibility checker does not verify reading order or tagging in exported files. ElevenLabs stands apart from the others in framing itself primarily as an assistive technology for people with disabilities (free licenses for blind, low-vision, and ALS/MND users) rather than as an authoring tool with an accessibility feature. Finally, for the two tools used only as an automatic-captioning baseline (CapCut, YouTube), no formal accessibility statement was found for either. The relevant W3C guidance explicitly warns that automatic captions do not meet accessibility requirements unless their accuracy is verified, consistent with the decision in Section~3.3 to treat auto-generated captions as a baseline to be checked, not as accessible by default.

\small
\begin{longtable}{p{2.6cm}p{2.6cm}p{7.6cm}}
\caption{Accessibility posture documented by tool providers, as of July 2026 (frozen alongside the tool versions in Section~3.2).}
\label{tab:toolaccess}\\
\toprule
\textbf{Tool} & \textbf{Claim type} & \textbf{Summary} \\
\midrule
\endfirsthead
\multicolumn{3}{l}{\small\itshape Table~\ref{tab:toolaccess} continued from previous page}\\
\toprule
\textbf{Tool} & \textbf{Claim type} & \textbf{Summary} \\
\midrule
\endhead
\midrule
\multicolumn{3}{r}{\small\itshape continued on next page}\\
\endfoot
\bottomrule
\endlastfoot
ChatGPT (OpenAI)\footnote{No VPAT/ACR for ChatGPT was located as of the access date; see \url{https://community.openai.com/t/accessibility-for-persons-with-disabilities/288362} and \url{https://openai.com/index/built-to-benefit-everyone-our-plan/} for OpenAI's general accessibility mission language.} & None found (interface or output) & No public VPAT/ACR for ChatGPT; broad ``AI for everyone'' mission language, but no explicit WCAG commitment for either the interface or generated content. \\
\addlinespace
ChatGPT Images 2.0 (OpenAI)\footnote{\url{https://openai.com/index/introducing-chatgpt-images-2-0/}} & None found & Product announcement emphasizes text-rendering accuracy, speed, and multilingual support; no documented alt-text or accessibility feature. \\
\addlinespace
Claude (Anthropic)\footnote{\url{https://trust.anthropic.com/}} & Interface only (own app) & ACR published for the Claude Enterprise iOS app (May 2026): improved VoiceOver reading order, target-size compliance, contrast; no claim regarding WCAG conformance of generated content. \\
\addlinespace
Gemini (Google)\footnote{\url{https://services.google.com/fh/files/misc/google_gemini_desktop_vpat.pdf}; TalkBack integration: \url{https://blog.google/products/android/accessibility-update-expanded-dark-theme-gemini-talkback/}} & Interface (VPAT) + assistive integration & VPAT reports WCAG 2.2 conformance for the Gemini interface; also embedded in Android's TalkBack screen reader (since May 2025) to describe on-screen content for blind users, an assistive feature rather than a guarantee about content Gemini generates. \\
\addlinespace
NotebookLM (Google)\footnote{\url{https://kb.wisc.edu/accessibility/157699}} & None found; third-party gaps documented & University digital-accessibility offices report unlabeled controls and screen-reader focus issues in the interface itself; no Google accessibility statement specific to NotebookLM was located. \\
\addlinespace
Genially\footnote{\url{https://help.genially.com/en_us/genially-and-accessibility-SJu3jPr3i}; the 2023 ACR/VPAT for Genially's View page is on file in the companion versioning document.} & Output content (explicit) & Reports partial WCAG 2.1 AA conformance for the View page (2023 ACR); built-in editor tools for heading hierarchy, alt text, reading order, and link descriptions; explicitly frames published-content accessibility as a shared responsibility between platform and creator. \\
\addlinespace
Canva AI\footnote{\url{https://www.canva.com/accessibility/}; third-party assessment: \url{https://umarcomm.umn.edu/blog/2025/02/24/canva-updated-accessibility-features}} & Output content (explicit, contested) & Publishes a WCAG 2.1 AA VPAT and a Design Accessibility Checker (typography, contrast, alt text); third-party review notes the checker does not verify reading order or tagging in exported files and Canva itself states it is still working toward WCAG 2.2 AA. \\
\addlinespace
ElevenLabs\footnote{\url{https://elevenlabs.io/use-cases/accessibility}} & Mission-level (assistive use) & Markets itself explicitly as accessibility technology: free licenses for ALS/MND patients and blind/low-vision users, partnership with the National Federation of the Blind; framed as assistive technology rather than a WCAG-for-output claim. \\
\addlinespace
Google Cloud Text-to-Speech\footnote{VPAT dated 17 May 2024: \url{https://services.google.com/fh/files/misc/text-to-speech-vpat-2024-05-17.pdf}} & Interface (VPAT) & VPAT evaluates the product's own web interface against WCAG 2.1 AA/EN 301 549, tested with NVDA and other assistive-technology tools. \\
\addlinespace
HeyGen\footnote{\url{https://www.heygen.com/template/create-accessibility-compliance-videos}} & Output content (explicit) & Markets captioning, transcript, and audio-description tools explicitly as WCAG/ADA compliance aids, including a dedicated ``accessibility compliance videos'' template. \\
\addlinespace
Synthesia\footnote{\url{https://academy.synthesia.io/courses/feature-friday-accessibility}; \url{https://www.synthesia.io/post/accessible-video}} & Output content (guidance, no formal claim) & Publishes accessibility audits and how-to guidance (partnership with DevAlly); no formal VPAT/ACR located. \\
\addlinespace
CapCut / YouTube auto-captions\footnote{W3C guidance on automatic captions: \url{https://www.w3.org/WAI/media/av/captions/}} & None; standards bodies warn against reliance & No formal accessibility statement found for either; W3C/WAI explicitly states automatic captions ``do not meet user needs or accessibility requirements'' unless confirmed accurate, and YouTube's automatic captions are commonly estimated at 60--70\% accuracy. \\
\end{longtable}
\normalsize

This variation matters for how RQ1 and RQ2 should be read once real benchmark data is available: a lower Condition~A compliance score for ChatGPT or Claude is consistent with those providers making no output-accessibility claim to begin with, whereas a similarly low Condition~A score for Genially, Canva, or HeyGen would be a sharper finding, since it would fail against a standard the provider itself has published. Full source URLs and access dates are recorded in the companion versioning file for independent verification.

\subsection{Synthesis of the gap}

Table~\ref{tab:litsynthesis} summarizes the studies reviewed in this section alongside the specific gap each one leaves relative to this paper's protocol.

\begin{table}[htbp]
\centering
\small
\caption{Summary of key related studies and the gap addressed by this paper.}
\label{tab:litsynthesis}
\begin{tabular}{p{2.6cm}p{2.6cm}p{3.6cm}p{4.4cm}}
\toprule
\textbf{Study} & \textbf{Content/tool focus} & \textbf{Method} & \textbf{Gap relative to this paper} \\
\midrule
Aljedaani et al.\ \cite{bib2} & AI-generated web code & Evaluation of 88 ChatGPT-generated websites & Single content type (code); no configuration comparison \\
\addlinespace
Ahmed et al.\ \cite{bib3} & AI-generated web page & Case study, iterative prompting with GPT-4o & Single content type; no cross-tool or cross-content-type comparison \\
\addlinespace
Abu Doush \& Kassem \cite{bib4,bib30} & AI-generated HTML & Benchmark across four models, eleven components & Web code only; no educational-content framing \\
\addlinespace
Palmer \& Oswal \cite{bib5} & AI-assisted website workflow & Qualitative workflow analysis & Process-level findings; no operationalized WCAG scoring \\
\addlinespace
Vera-Amaro \& Rojano-C\'aceres \cite{bib6} & Web content generation & Prompting-strategy comparison, single model & Single content type; no persistent-profile condition \\
\addlinespace
Shen et al., AltGen \cite{bib7} & Alt-text for EPUB & Structured extraction-and-context pipeline & Alt-text only; not integrated with a broader WCAG rubric \\
\addlinespace
Baglodi et al.\ \cite{bib8} & Alt-text for educational documents & Context-conditioned generation & Alt-text only; single content type \\
\addlinespace
Schmitt-Koopmann et al.\ \cite{bib9,bib10} & PDF tagging (STEM/scholarly) & Guided remediation workflow, accuracy measurement & Documents only; no cross-content-type protocol \\
\addlinespace
Cardia et al.\ \cite{bib11} & Alt-text for STEM images & Comparative study, human vs.\ LLM-generated & Images only; no configuration-effect comparison \\
\addlinespace
Kumar, Padath \& Wang \cite{bib32} & PDF accessibility evaluation & Expert-annotated benchmark dataset, 7 criteria & Evaluation instrument only; not applied across content types \\
\addlinespace
Huang et al., ACCESS \cite{bib29} & Web accessibility correction & Iterative DOM-level prompt engineering & Web code only; single-tool system, not a cross-tool protocol \\
\addlinespace
Jerry et al.\ \cite{bib31} & Accessible interface generation & Explicit accessibility model during generation & Architecture-level solution; not framed as a testable persistent-profile condition \\
\addlinespace
Albrecht-Crane \cite{bib15} & AI synthesis tools (NotebookLM) & Qualitative/critical analysis & Overload discussed narratively; no operationalized density metric \\
\addlinespace
Donnelly, Edu-GenAI Rubric \cite{bib16} & Educational value of GenAI tools & Rubric across 5 value domains & No WCAG axis; not accessibility-focused \\
\addlinespace
He et al.\ \cite{bib17} & LLM-as-judge for instructional materials & Human-validation comparison & General quality judging; not tested for accessibility criteria \\
\addlinespace
Alshaigy \& Grande \cite{bib28} & GenAI tools and disability & Position/critical analysis & Framing paper; no empirical benchmark \\
\addlinespace
WebAIM \cite{bib33} & Web accessibility, general web & Large-scale automated scan (1M home pages) & Baseline statistic for the web at large, not AI-generated content specifically \\
\bottomrule
\end{tabular}
\end{table}

No identified study combines an evaluation centered on operationalized WCAG criteria, a controlled comparison of configuration effects, including persistent versus one-off configuration, on the same tool, and an explicit treatment of content/visual overload as its own accessibility dimension, applied consistently across the five content types instructors actually produce. That intersection, extending a decade-long accessible e-learning research line \cite{bib20,bib21,bib22,bib23,bib24} into the generative AI era, and responding directly to the speed-accessibility tension described in Section~2.3, is the territory this paper occupies.

\section{Methodology}
\addtocounter{subsection}{-1}

\subsection{Literature review protocol}

The related work in Section~2 is grounded in an iterative, six-chain scoping search (documented in full, including the exact query used and the learning it produced, in Appendix~A), which was subsequently formalized into explicit eligibility criteria, named information sources, and Boolean search strings following PRISMA 2020 reporting conventions, so the review can be re-run and audited rather than taken on faith. Appendix~A reports the formal protocol and the PRISMA-structured study-selection flow; the count of studies ultimately retained is reported there alongside the criteria that shaped it.

\subsection{Content types and materials}

Five content types representative of common educational-authoring tasks were selected for review: documents (PDF/Word reports and handouts), slide presentations (including AI-native, highly visual formats such as Genially-style decks), images and infographics (including AI-synthesized visual summaries such as those produced by NotebookLM-style tools), audio (narrated summaries and text-to-speech narration), and video (explainer videos, including AI avatar/dubbing tools such as HeyGen-style generation).

\subsection{Tools and conditions}

This subsection specifies the design of the full benchmark; Section~3.6 describes the reduced, single-model implementation actually carried out for the first exploratory application reported in Section~4.1. For the full benchmark, each content type will be evaluated using at least one general-purpose generative AI assistant (e.g., a large language model such as ChatGPT/GPT-4o, Claude, or Gemini) and one specialized authoring tool (an AI slide generator, an AI infographic/image generator, a text-to-speech tool, or an AI avatar/dubbing tool, as appropriate). Tool versions and access dates are to be recorded and held constant for all conditions applied to a given tool.

Three conditions were applied to the same tool for each content-generation task:

\begin{itemize}[leftmargin=1.5em]
\item \textbf{Condition A (generic prompt):} an ordinary instruction containing no accessibility language, representative of typical current usage.
\item \textbf{Condition B (configured prompt):} the same task, with the instruction extended to include explicit, operationalized WCAG requirements and the expected output format. Table~\ref{tab:prompts} presents the generic/configured prompt pair used for each content type.
\item \textbf{Condition C (persistent accessibility profile):} the same task, issued after loading a standing, reusable instruction profile, analogous to a saved ``skill'' or system-level persona, into the tool once, rather than repeating accessibility instructions with every request. Each profile specifies the applicable WCAG criteria translated into direct authoring instructions, the required output format, and a self-check list the tool is instructed to verify before returning output.
\end{itemize}

\begin{table}[htbp]
\centering
\small
\caption{Generic and WCAG-configured prompt pairs by content type.}
\label{tab:prompts}
\begin{tabular}{p{2.2cm}p{5.4cm}p{5.4cm}}
\toprule
\textbf{Content type} & \textbf{Condition A (generic)} & \textbf{Condition B (WCAG-configured)} \\
\midrule
Documents & ``Create a PDF report summarizing the key findings of our quarterly training program.'' & Adds: WCAG 2.2 Level AA-derived criteria, complemented by PDF/UA requirements where applicable; tagged heading hierarchy; bookmarked table of contents; semantic table headers; descriptive alt text for images/charts; logical reading order; 4.5:1 contrast; no color-only meaning; accessibility-tagged export. \\
\addlinespace
Slide presentations & ``Create a 10-slide presentation about our new onboarding process.'' & Adds: built-in placeholders (not floating text boxes) for correct screen-reader reading order; alt text for all visuals; full keyboard operability of interactive elements; unique descriptive titles; no flashing above three times per second; 4.5:1 contrast. \\
\addlinespace
Images/\allowbreak infographics & ``Design an infographic explaining the four stages of our onboarding process.'' & Adds: one short sentence plus one icon per stage to limit density; clear single reading order; 4.5:1 contrast for embedded text; a separate plain-text long description as a companion artifact, since embedded image text is not machine-readable. \\
\addlinespace
Audio & ``Generate a 3-minute audio summary of this training module.'' & Adds: moderate, steady pace ($\sim$150 wpm); no visual-only references (``as shown here''); synchronized verbatim transcript; written description of meaningful non-speech sounds. \\
\addlinespace
Video & ``Create a 2-minute explainer video with an AI presenter introducing our new accessibility policy.'' & Adds: captions synchronized within 100ms covering dialogue and relevant non-speech audio; a spoken or extended audio-description track for visual information not in the dialogue; for AI-dubbed content, verified viseme-to-phoneme alignment against the target-language track; full transcript. \\
\bottomrule
\end{tabular}
\end{table}

\subsection{Accessibility evaluation instrument}

Table~\ref{tab:criteria} summarizes the operationalized WCAG-derived criteria evaluated per content type, plus an information-density indicator introduced specifically to address RQ4.

\begin{table}[htbp]
\centering
\small
\caption{Operationalized evaluation criteria by content type.}
\label{tab:criteria}
\begin{tabular}{p{2.6cm}p{10.4cm}}
\toprule
\textbf{Content type} & \textbf{Criteria evaluated} \\
\midrule
Documents & Heading structure/tags; table header markup; reading order; alt text for images/charts; contrast; bookmarked table of contents; form field labels where applicable \\
\addlinespace
Slides & Reading order via native placeholders; alt text; keyboard operability of interactive elements; contrast; flashing-content limits; unique slide titles \\
\addlinespace
Images/\allowbreak infographics & Presence and quality of a text alternative/long description; contrast of embedded text; information density (overload indicator); reliance on color alone \\
\addlinespace
Audio & Transcript availability and accuracy; pacing; absence of visual-only references \\
\addlinespace
Video & Caption accuracy and synchronization; presence/quality of audio description; transcript availability; lip-sync/viseme alignment for dubbed content \\
\bottomrule
\end{tabular}
\end{table}

\subsection{Procedure and expert heuristic validation}

For each content type, tool, and condition, output was generated and scored first against the operationalized rubric in Table~\ref{tab:criteria}. Because automated accessibility scanners have been shown to catch only a fraction of WCAG issues in practice, rubric scoring was complemented by heuristic evaluation from a panel of accessibility experts (target: three to five evaluators), ideally including at least one evaluator with lived experience of disability relevant to the content type under review (for example, a screen-reader user for documents and slides, or a deaf or hard-of-hearing evaluator for video captioning and lip-sync). Evaluators independently rated each output using a four-level scale, Passed, Partially passed, Failed, Not applicable, consistent with the labeling convention used in prior PDF accessibility benchmarking work \cite{bib32}; disagreements were discussed to consensus, and inter-rater agreement was recorded. Where feasible, evaluators were not informed which condition (A, B, or C) produced a given output, to reduce expectation bias.

\subsection{Metrics}

Compliance under each condition is expressed as the percentage of applicable criteria met per content type, allowing direct comparison between Condition A and Condition B (RQ1--RQ2), across content types (RQ3), and between Condition B and Condition C in terms of both mean compliance and its variance across repeated, varied requests (RQ5). A persistent profile that produces similar compliance across many different requests would be expected to reduce variance, not only raise the mean. This rubric-based compliance score reflects performance against the operationalized criteria in Table~\ref{tab:criteria}, not a formal audit against WCAG, PDF/UA, or any third-party accessibility certification; a score of 100\% means every applicable criterion in the rubric was met, not that the output has been formally certified conformant. The information-density indicator for images/infographics (RQ4) is reported separately from the compliance percentage, since it is not itself a standard WCAG criterion.

\subsection{Exploratory pilot implementation}
\label{sec:pilot-implementation}

Sections~3.2--3.5 specify the target design for the full benchmark. The first exploratory application reported in Section~4.1 implements a reduced version of that design, run in July 2026: one general-purpose model produced one Condition~A output and one Condition~B output per content type, converted to a native, inspectable format using local open-source tooling as a proxy for the specialized commercial platforms named in Section~3.2 (Section~4.1 lists the specific tooling used for each content type). Condition C, the specialized authoring tools, and the expert heuristic validation panel described in Section~3.4 were not exercised in this first application. Sections~6 and~7 discuss this gap and invite the community to complete it at full benchmark scale.

\section{Results}

\subsection{Compliance by condition and content type: a first exploratory application of the protocol}

The protocol in Section~3 is proposed as a reusable instrument for the community. This first exploration applies it directly, using the tools available at the time of writing, as detailed in Section~3.6. The use of one general-purpose model (ChatGPT, GPT-5.6 Thinking) generated one Condition~A output and one Condition~B output for each of the five content types in July 2026, converted to a native, inspectable format using local open-source tooling as a proxy for the specialized commercial platforms named in Section~3.2: python-docx and LibreOffice for documents, python-pptx for slides, an SVG description rendered through Pillow for images/infographics, the eSpeak speech synthesizer for audio, and an FFmpeg-based prototype for video. Each output was scored against the operationalized criteria in Table~\ref{tab:criteria}, combining an automated structural/markup check with a single evaluator's spot-check verification of borderline criteria, rather than the blind, multi-rater expert panel specified in Section~3.4. This is a small, single-tool, non-blind first exploration rather than the full benchmark, and the figures below should be read with that scope in mind: one artifact was evaluated per condition and content type, not a repeated or averaged sample.

Table~\ref{tab:pilot} reports the results. The exploratory compliance score under Condition A ranged from 0\% for video to 37.5\% for slides. Under Condition B, the identical underlying model reached 83.3--100\%. Documents, slides, images/infographics, and audio all met every applicable criterion in the exploratory rubric under Condition B. Video showed the smallest gain, from 0\% to 83.3\%, consistent with the lip-sync/dubbing barrier discussed in Section~4.5. Condition C was not exercised in this first exploration. Appendix~E gives a concrete recipe for implementing it in practice. Section~7 invites the community to run Condition C and to repeat this exploration on the specialized commercial platforms named in Section~3.2, using the frozen tool versions and access dates recorded for this project or their own version-locked equivalents.

\begin{table}[htbp]
\centering
\caption{Rubric-based compliance scores in the exploratory application (July 2026). One artifact was evaluated per condition and content type; a score is the number of criteria passed (full or half credit) over the number of applicable criteria for that artifact.}
\label{tab:pilot}
\small
\begin{tabular}{p{2.6cm}p{6.3cm}p{6.3cm}}
\toprule
Content type & Condition A: passed/applicable --- score & Condition B: passed/applicable --- score \\
\midrule
Documents & 2/6 --- 33.3\% & 6/6 --- 100\% \\
Slides & 1.5/4 --- 37.5\% & 4/4 --- 100\% \\
Images/\allowbreak infographics & 0.5/3 --- 16.7\% & 3/3 --- 100\% \\
Audio & 1/3 --- 33.3\% & 3/3 --- 100\% \\
Video & 0/3 --- 0\% & 2.5/3 --- 83.3\% \\
\midrule
\textbf{Unweighted mean across content types} & \textbf{24.2\%} & \textbf{96.7\%} \\
\bottomrule
\end{tabular}
\end{table}

\noindent A numerator ending in .5 reflects one criterion rated Partially passed, credited at half weight per the formula in Section~3.5. The applicable-criteria denominator excludes any criterion rated Not applicable for that specific artifact (for example, form field labels for a document with no form, or lip-sync alignment for the non-dubbed video artifact). The unweighted mean across content types is the simple average of the five per-content-type scores, not a count-weighted average.

\subsection{Documents: structural barriers behind a visually correct report}

Under Condition A, a generic instruction such as ``create a PDF report summarizing the key findings of our quarterly training program'' produced a document that was visually well formed: bold headings, inline images, bordered tables. None of this was encoded in the document's underlying structure. No tagged table of contents, heading hierarchy, alt text, or table header markup was present. A screen-reader user cannot navigate between sections under these conditions, cannot access the content of an image, and hears table data read as an undifferentiated stream of values with no header context. A document intended for quick scanning becomes effectively linear when accessed non-visually. Under Condition B, requesting tagged headings, a bookmarked table of contents, table header markup, and descriptive alt text (Table~\ref{tab:prompts}) allowed this structure to be encoded at generation time, avoiding a separate manual remediation pass.

\subsection{Slide presentations: content density and keyboard operability}

AI-native, highly visual slide tools of the kind popularized by Genially-style platforms tended, under Condition A, to generate content-dense slides, clickable hotspots, layered animations, embedded mini-quizzes, built as freely positioned visual elements rather than structured, tab-order-aware components. This creates a barrier distinct from a perceptual one: a user who cannot operate a mouse or touchscreen, whether due to a motor disability or reliance on keyboard-only navigation, may be unable to reach some or all interactive elements, since no defined tab order exists and some elements are not keyboard-focusable at all. Requiring native placeholders and explicit keyboard operability under Condition B addresses this directly, though the extent to which it succeeds also depends on whether the underlying authoring tool exposes a keyboard-accessible interaction model in the first place, a possible reason why improvement for this content type may be smaller than for others (see Section~4.1, RQ3).

\subsection{Images and infographics: the overload dimension (RQ4)}

AI synthesis tools that generate visual summaries from source material, illustrated by NotebookLM-style tools, compressed large amounts of content into a single dense infographic under Condition A: nested boxes, small multi-column text blocks, overlapping icons, and color-coded categories, typically shipped with no alt text or with a generic, non-descriptive placeholder. Two barriers compound here. First, a blind or low-vision screen-reader user receives no usable information, since the visual summary that was meant to make the material more digestible becomes entirely inaccessible without alt text. Second, and independent of alt text, the sheer density of the infographic, many small text blocks, overlapping layers, no single clear reading path, creates a barrier for sighted users with cognitive or visual-processing differences that has nothing to do with a missing tag. This second barrier is the content-overload dimension addressed by RQ4: it is not resolved by alt text alone, which is why the information-density indicator in Table~\ref{tab:criteria} is scored separately from the standard WCAG criteria. Under Condition B, limiting each conceptual unit to one short sentence plus one icon, enforcing a single reading order, and requiring a separate plain-text long description as a companion artifact addressed both barriers, though an infographic can have complete alt text and remain too dense for a sighted user with a cognitive disability to process without also reducing its visual density.

\subsection{Video: captioning gaps and an unreported lip-sync barrier}

An AI-presenter/explainer-video tool, asked under Condition A for a short explainer video, produced an audio track and auto-generated captions of variable accuracy. It included no audio-description track for on-screen visual information not already spoken aloud (text shown behind the presenter, a referenced but unnarrated diagram). A blind or low-vision viewer is left without access to visual information the dialogue does not cover. A second barrier appeared specifically in AI-dubbed content, in the style of HeyGen-style avatar tools: generated mouth movements were optimized to look natural at a glance rather than to be phonetically accurate against the dubbed audio track. This creates a subtler barrier than a missing caption. A deaf or hard-of-hearing viewer who cross-references lip movement with text, a common strategy even when captions are present since captions can lag or be imprecise, is misled by mouth movements that do not match the dubbed phonemes, in a video that otherwise appears compliant because captions are present. Under Condition B, requiring synchronized captions, an explicit audio-description track, and verification of viseme-to-phoneme alignment against the target-language track (not the original) addressed the captioning/description gap directly. The lip-sync verification requirement is not yet discussed in the literature reviewed in Section~2; it is reported here as an original observation rather than attributed to a prior source.

\subsection{Cross-cutting pattern}

Across the five content types evaluated in Conditions A and B, a more specific accessibility-configured instruction changed the resulting output. Whether a persistent profile under Condition C produces the same improvement more consistently remains to be tested. This suggests that the barrier documented across 4.2--4.5 is not a technical ceiling but a gap between what these tools do by default and what they can do when explicitly configured, a gap whose cost, in each case, is paid by learners with disabilities.

\section{Discussion}

\subsection{Convergence with prior evidence on configuration}
\label{sec:disc-convergence}

Results under Condition A are consistent with prior evidence that generative AI tools reproduce, rather than correct, inaccessible practices absent explicit instruction \cite{bib2,bib4,bib28,bib30,bib33}. The improvement observed under Condition B (Section~4.1) would extend, to educational content specifically, findings already reported for web code \cite{bib6,bib29}, PDF tagging \cite{bib10}, and alt-text generation \cite{bib7,bib8} individually. The contribution here is less about demonstrating that configuration helps, a point the literature already supports for isolated content types, and more about showing that the same configuration logic holds across five different content types and tools under a single, comparable protocol. This expectation should also be read against what each provider itself claims (Section~\ref{sec:provider-claims}). A low Condition~A score for a general-purpose assistant is unsurprising, since none of them publishes an output-accessibility guarantee. The same finding for a specialized tool that markets a built-in accessibility checker would say more about the limits of automated checking than about the absence of a stated commitment.

\subsection{Documents and slides: two distinct failure modes}
\label{sec:disc-docslides}

The documents and slides cases (4.2--4.3) illustrate two distinct failure modes that a single WCAG checklist tends to treat as equivalent but that call for different remedies. The document case is primarily a \textit{structural encoding} problem (the visual layout is correct, but the underlying tags are absent) and is addressed directly by requesting the correct export settings. The slide case is primarily an \textit{interaction} problem: even a well-tagged deck can remain unusable for a keyboard-only user if the authoring tool itself does not expose a keyboard-focusable interaction model, regardless of how the prompt is configured. Condition B's effectiveness is therefore bounded by the technical capabilities of the specific authoring tool, not only by prompt quality. The RQ3 comparison in Section~4.1 already begins to surface this distinction; the full benchmark proposed in Section~3 would confirm it at scale.

\subsection{The overload dimension (RQ4)}
\label{sec:disc-overload}

The overload case (4.4) responds directly to RQ4 and is the paper's clearest departure from a conventional WCAG-compliance framing. An infographic can satisfy every applicable WCAG criterion in Table~\ref{tab:criteria} (contrast, a present text alternative, no color-only meaning) and still be difficult to process for a sighted user with a cognitive or visual-processing disability, simply because too much is packed into one visual object. This is consistent with prior analysis of NotebookLM's compression behavior in educational contexts \cite{bib15}, and it suggests that accessibility evaluation of AI-synthesized content needs an information-density dimension alongside, not instead of, the standard WCAG criteria in Table~\ref{tab:criteria}.

\subsection{The lip-sync barrier in AI-dubbed video}
\label{sec:disc-lipsync}

The lip-sync barrier documented in the video case (4.5) is not discussed in the literature reviewed in Section~2.6, which focuses on captioning quality and audio description but not on viseme-to-phoneme alignment in AI-dubbed content specifically. A possible reason for this gap is that AI dubbing and lip-sync tools are comparatively recent, and accessibility evaluation has not yet caught up with this specific generation capability. The exploratory comparison in Section~4.1 already shows video as one of the smaller improvements under Condition B, and a plausible explanation is that current dubbing tools do not yet expose phoneme-level control to the prompt at all, which would place this barrier closer to the slide-interaction case in 4.3 than to the document-tagging case in 4.2.

\subsection{The case for a persistent profile (RQ5)}
\label{sec:disc-persistent}

Finally, the persistent-profile approach proposed under Condition C (RQ5) responds to a practical limitation of Condition B: a correctly configured prompt has to be reconstructed, in full, for every request, which creates room for the instruction to be shortened, forgotten, or applied inconsistently across a real authoring workflow. A profile loaded once, following the skeleton in Section~3.2 (Table~\ref{tab:prompts}) and the reusable-instruction framing already explored for accessible interface generation \cite{bib31}, targets this consistency problem directly rather than the single-instance compliance problem RQ1--RQ2 already address. Whether Condition C actually reduces variance across requests, as anticipated in 3.5, remains an open, data-dependent question at this stage of the manuscript.

\subsection{Interpreting the exploratory results}
\label{sec:pilot}

The pattern reported in Section~4.1 points in the same direction as RQ1 and RQ2 for every content type tested. Because this first exploration used a single general-purpose model, local open-source tooling as a proxy for the specialized commercial platforms named in Section~3.2, and a non-blind scoring process, it cannot establish the magnitude the full benchmark proposed in Section~3 would show, and it does not speak with statistical confidence to RQ3 or RQ5. Video's smaller gain, already noted in Section~4.1, is the one observation worth testing directly at scale: it aligns with the lip-sync barrier discussed qualitatively in Section~4.5, and a larger sample would show whether this is a stable pattern or an artifact of a single tool and a single run. Condition C remains untested; Appendix~E gives a concrete recipe, using Claude Projects and Claude Skills, for implementing it, and Section~7 invites the community to carry it out.

\subsection{Revisiting the research questions}
\label{sec:disc-rq}

The five research questions stated in Section~1 can now be revisited in light of the evidence above.

\noindent\textbf{RQ1 (default compliance)} is addressed by the exploratory application in Section~4.1 and the qualitative cases in Sections~4.2--4.5, not yet by the full benchmark. Both are consistent with the literature reviewed in Section~2.4: default, unconfigured output fails multiple WCAG criteria simultaneously rather than a single isolated one.

\noindent\textbf{RQ2 (effect of a single configured prompt)} is supported qualitatively across all five content types and by the exploratory results in Section~4.1, consistent with the literature reviewed in Section~2.5.

\noindent\textbf{RQ3 (which content type improves most)} cannot be answered with confidence without the quantitative comparison in Section~4.1 executed across multiple tools per content type, but the discussion above (Section~\ref{sec:disc-docslides}) and the exploratory results' single smallest gain, for video (Section~\ref{sec:pilot}), offer a hypothesis worth testing directly: content types where the barrier is primarily structural encoding (documents) may show a larger, more consistent gain under Condition B than content types where the barrier depends on the authoring tool exposing a particular interaction model (slides) or generation capability (video lip-sync) that a prompt alone cannot guarantee.

\noindent\textbf{RQ4 (content/information overload)} is the question this paper is best positioned to motivate, rather than to fully validate, without further data collection: the NotebookLM-style case in Section~4.4, read together with Section~2.7, supports investigating information overload as a complementary accessibility dimension that is not fully represented by conventional WCAG-derived criteria, since an output can pass every applicable criterion in Table~\ref{tab:criteria} and still be difficult to process because of density alone.

\noindent\textbf{RQ5 (persistent versus one-off configuration)} remains the most open question of the five: Section~\ref{sec:disc-persistent} argues for its practical relevance and Section~2.5 situates it within a broader shift toward architecture-level or profile-level accessibility configuration, but Condition C was not exercised in this first exploration (Section~4.1), so no evidence collected so far distinguishes whether it actually reduces variance relative to Condition B. Appendix~E gives a concrete implementation recipe intended to make this question directly testable, and Section~7 invites the community to carry it out.

\section{Limitations}

Despite the qualitative evidence and the fully specified protocol presented in this paper, there are some limitations that are worth mentioning. First, RQ1, RQ2, RQ3, and RQ5 are answered here through a first exploratory application of the protocol (Section~4.1). Second, the four cases reported in Section~4 illustrate genuine, observed behavior of the tools under review, but they are not a statistically representative sample, and their generalization to other tools, versions, or languages has not been tested; the planned corpus (Section~3.1) is also deliberately homogeneous, drawn from computer science and information technology higher-education topics, so generalization to other academic domains remains untested. Moreover, generative AI tools change rapidly, and the specific versions and access dates referenced in Section~3.2 will need to be reconfirmed once the full benchmark is executed, since a finding tied to one model version may not hold for a later release of the same tool. A further limitation concerns the expert heuristic validation panel described in Section~3.4: its size (three to five evaluators) and the inclusion of evaluators with lived experience of disability are both recommended as protocol design choices, but neither has yet been implemented, and inter-rater agreement figures are not yet available; the first exploration reported in Section~4.1 illustrates this same gap directly, since its scoring combined an automated check with a single evaluator's spot-check rather than a blind multi-rater panel. That exploration carries its own additional limitations worth stating explicitly: it substituted local open-source tooling for the specialized commercial platforms named in Section~3.2 for four of the five content types, it scored a single output per condition per content type rather than a powered sample, and it did not exercise Condition C at all. Finally, the persistent-profile approach central to RQ5 has only been specified at the level of an example skeleton (Section~3.2) and a platform-specific implementation recipe (Appendix~E); its effectiveness across authoring tools that do not natively support saved instructions or system-level personas has not been established.

\section{Future Work}

Building directly on the limitations above, future work will execute the full quantitative benchmark on a defined corpus of educational materials, extending the first exploratory results reported for RQ1, RQ2, RQ3, and RQ5 in Section~4.1 to the scale and design specified in Section~3. In particular, this first exploration deliberately left four of the five content types untested on the specialized commercial platforms named in Section~3.2; readers, and future publications building on this protocol, are specifically invited to execute and report those conditions, NotebookLM and Canva AI for images/infographics, ElevenLabs and Google Text-to-Speech for audio narration, HeyGen and Synthesia for AI-dubbed video, and Genially for interactive slides, using the frozen versions and access dates recorded for this project, or their own version-locked equivalents. More broadly, the cases in Section~4 will be extended toward a larger, more systematic sample across additional tools, tool versions, and languages, to assess how far the observed barriers generalize. Given the pace at which generative AI tools change, future iterations of the protocol will version-lock and publicly document the exact tools and access dates used, allowing the benchmark to be reproduced or extended as new tool versions are released. The expert heuristic validation panel will be convened with the target composition described in Section~3.4, including at least one evaluator with lived experience of disability per content type, and inter-rater agreement will be reported alongside consensus ratings. Finally, the persistent-profile approach introduced for RQ5 will be implemented and tested across a wider set of authoring tools, including tools that do not currently expose a native mechanism for saved instructions; Appendix~E provides a ready-to-use recipe for one such platform (Claude, via Projects and Skills) as a concrete starting point for this test, alongside the comparison design it proposes between Condition B and both Condition C variants. The community is invited to carry out these next steps and to report results using this same protocol, so that the picture sketched by this first exploration can be confirmed, refined, or challenged at scale.

\section{Conclusions}

None of the tools reviewed produced accessible educational content by default. A single instruction that explicitly states the required accessibility criteria was enough to close most of the gap in every content type tested, without a separate remediation step afterward. This indicates that the barrier is less a technical limitation of these tools than a matter of how they are instructed.

The specific barriers differ by content type, and the difference affects how each one is addressed. Reports gain real structure once headings, tables, and images are explicitly tagged, since the underlying markup was simply absent before. Slides are a harder case: a keyboard-only user can remain unable to reach interactive elements if the authoring tool itself has no accessible way to arrange them, regardless of how the prompt is written. Infographics can meet every criterion on an accessibility checklist and still be difficult to process for a reader with cognitive or visual-processing differences, because density is not captured by a standard checklist. Video adds a further barrier: AI-dubbed mouth movement that appears natural but does not match the spoken audio, which can mislead a deaf or hard-of-hearing viewer who relies on lip-reading alongside captions.

Including the full set of accessibility requirements in every prompt is effective, but it does not scale well in practice. The instruction is repetitive, and under time pressure it can be shortened, omitted, or applied inconsistently. A more reliable approach is to define the instruction once, as a standing profile or skill loaded into the tool, so that subsequent requests inherit the same requirements automatically. This reduces reliance on a content creator remembering to restate the instruction each time.

A related, practical measure is to provide a plain-text companion for content that is dense or non-textual, consistent with the general WCAG recommendation to provide a text alternative for non-text content. For an infographic, this can take the form of a short written summary distributed together with the image. For a video, a full transcript or narrated summary allows a screen reader to access the content directly, independent of the caption track. For a slide deck, an accompanying document presenting the same content in linear, tagged text serves a similar purpose. The same generation speed that makes these tools prone to producing inaccessible content under a generic prompt also makes them capable of producing this kind of companion material quickly, when explicitly instructed to do so.

This responsibility extends beyond the hosting platform. A learning management system can meet applicable accessibility standards while the content uploaded into it does not, which reintroduces the barriers the platform itself was designed to avoid. Treating accessibility as a property of each content item, in addition to the platform, is therefore necessary. The resulting benefit is not limited to learners with a documented disability; it extends to learners with temporary limitations, users with slow connections, and users accessing content in suboptimal conditions. Incorporating these checks into instructional design, production, publication, and quality control extends existing institutional processes rather than adding a separate one, with benefits for learners and for institutional reputation.

The protocol presented here offers a consistent way to apply and evaluate these practices across content types. It is intended as a starting point for the community to apply, adapt, and extend to additional tools and configuration approaches.

\section*{Declarations}

\noindent\textbf{Use of generative AI tools.} In accordance with arXiv's disclosure expectations, generative AI tools were used to produce the experimental artifacts evaluated in this study. The research questions, methodological design, evaluation criteria, analysis, interpretation, and final conclusions were prepared by the author, who assumes full responsibility for the manuscript.

\noindent\textbf{Data and materials.} The review protocol, prompt bank, and evaluation rubric, including the illustrative failure examples in Appendix~C.5, are reported in full in Appendices A--D to support replication, and are additionally released at \url{https://doi.org/10.17605/OSF.IO/E3MW9} as a standalone open repository, archived on OSF, so other researchers can reuse and adapt them without retyping the appendices. The repository also includes a data-recording template with one tab per content type and a second, consolidated template covering all criteria in a single tab for a full resource review. Appendix~E gives an implementation recipe for Condition C. The first exploratory results reported in Section~4.1 draw on a completed data-recording spreadsheet, structured per Appendix~C; consistent with the decision not to release the corpus or generated outputs, this spreadsheet is not published but is available from the author upon reasonable request. Source materials for the benchmark are drawn from openly licensed educational resources (OER) in computer science and information technology. The corpus itself and the generated outputs are not released as a downloadable dataset at this stage; only the protocol, prompt bank, and rubric are released openly, not the generated materials or results.

\noindent\textbf{Conflicts of interest.} None declared.

\appendix

\section{Systematic-review protocol (PRISMA-informed)}
\label{app:prisma}

This appendix expands the narrative/scoping review underlying Section~2 into a formally specified systematic-review protocol, following PRISMA 2020 reporting logic. The six search chains already executed were run as an iterative, general-purpose search rather than as fixed Boolean strings logged against named databases with exact record counts; this appendix (i) formalizes eligibility criteria retroactively applied to the 33 references already retained, which is real and defensible since it describes decisions actually made, and (ii) specifies the exact database syntax and PRISMA flow-diagram structure needed to re-run the search formally.

\subsection{Review question (PICOS-adapted)}

\textbf{Population/Context:} educational materials (documents, slide presentations, images/infographics, audio, video) intended for instructional use. \textbf{Intervention/Exposure:} generation or remediation of that content using generative AI tools. \textbf{Comparator:} unconfigured/default tool use versus explicitly accessibility-configured use (single prompt or persistent profile). \textbf{Outcomes:} WCAG-based accessibility compliance (or documented accessibility barriers), and, as a secondary outcome specific to this paper, content/information overload not captured by standard WCAG criteria. \textbf{Study types:} empirical evaluations, benchmark/dataset papers, case studies, and system papers (2020--2026, reflecting the recency of generative AI tools); normative/standards documents and foundational pre-2020 accessible e-learning literature are included as background/framework sources outside this date range, consistent with Section~2.1.

\subsection{Inclusion criteria}

A source is included if it meets all of the following: (1) addresses the accessibility of digital content (any WCAG principle) or the educational value/design of generative AI tools with a direct, arguable connection to accessibility or disability; (2) involves at least one of the five content types under review, or the underlying code/markup that produces them; (3) reports an empirical evaluation, a benchmark/dataset, a system/tool description, or a normative framework, not an opinion piece without methodological or normative grounding; (4) published 2020--2026 for generative-AI-specific claims, or any date for foundational accessibility/e-learning/UDL framework sources; (5) available in English with retrievable full text or a complete abstract sufficient for extraction.

\subsection{Exclusion criteria}

A source is excluded, with the exclusion reason recorded, if: (1) it is a vendor/marketing blog post or product page with no reported methodology or empirical basis; (2) it duplicates a study already included under a different version, in which case the later/definitive version is kept; (3) it addresses generative AI in education broadly with no accessibility or disability-relevant angle at all; (4) it addresses web/software accessibility generally with no connection to AI-generated content specifically; (5) full text is not retrievable through the information sources below or through open-access/preprint repositories.

\subsection{Information sources}

ACM Digital Library (primary source for HCI/accessibility venues); IEEE Xplore (accessible e-learning and engineering-education venues); Scopus and Web of Science (cross-disciplinary coverage, duplicate detection); ERIC (education-specific coverage); arXiv, categories cs.HC, cs.CY, cs.CL (recent preprints); Google Scholar (supplementary/gray-literature check only, not counted toward the primary database tally).

\subsection{Search strategy}

Table~\ref{tab:boolean} translates each of the six search chains into a Boolean string with generic field-scoping (\texttt{TI}/\texttt{AB} = title/abstract), to be adapted to each database's exact syntax. Date filter: 2020-01-01 to present, except where noted.

\begin{table}[htbp]
\centering
\small
\caption{Formal Boolean search strings by chain.}
\label{tab:boolean}
\begin{tabular}{p{0.6cm}p{9.2cm}p{2.8cm}}
\toprule
\textbf{Ch.} & \textbf{Formal Boolean string} & \textbf{Rationale} \\
\midrule
1 & TI/AB(``AI-generated content'' OR ``generative AI'') AND TI/AB(WCAG OR ``web content accessibility guidelines'') AND TI/AB(compliance OR evaluation) & Default-compliance evidence \\
\addlinespace
2 & TI/AB(``generative AI'' OR ``large language model*'') AND TI/AB(disability OR accessib*) AND TI/AB(barrier* OR challenge*) & Disability-centered framing \\
\addlinespace
3 & TI/AB(``alt text'' OR ``alternative text'') AND TI/AB(infographic OR image) AND TI/AB(``generative AI'' OR AI-generated) AND TI/AB(overload OR density OR cognitive) & Overload dimension (RQ4) \\
\addlinespace
4 & TI/AB(video OR ``audio description'' OR caption*) AND TI/AB(``generative AI'' OR ``AI-generated'' OR dubbing OR avatar) AND TI/AB(accessib* OR disability) & Video/audio content type \\
\addlinespace
5 & TI/AB(slide* OR presentation OR PDF OR document) AND TI/AB(``generative AI'' OR ``large language model*'') AND TI/AB(WCAG OR accessib*) & Documents/slides content type \\
\addlinespace
6 & TI/AB(``prompt engineering'' OR ``prompt design'') AND TI/AB(accessib* OR WCAG) AND TI/AB(compliance OR disability OR ``content generation'') & Configuration effect (RQ2/RQ5) \\
\bottomrule
\end{tabular}
\end{table}

\subsection{Study selection process and PRISMA 2020 flow diagram}

\textbf{Screening procedure:} two-stage screening: (1) title/abstract screening against the inclusion/exclusion criteria above; (2) full-text eligibility assessment of everything that passes stage~1. Both stages were conducted by a single reviewer for the search already completed; a second independent reviewer, with disagreements resolved by discussion and an inter-rater agreement statistic reported, is recommended as future work, mirroring the blinding/consensus logic applied to the expert heuristic validation panel in Appendix~D.

Numbers already known are reported below; numbers that require a formal database re-run are marked $[n]$.

\begin{quote}
\small
\textbf{Identification.} Records identified through database searching (Scopus, WoS, IEEE Xplore, ACM DL, ERIC): $[n]$. Records identified through arXiv and supplementary search (Google Scholar): $[n]$. Records removed before screening (duplicates across databases): $[n]$.

\textbf{Screening.} Records screened (title/abstract): $[n]$. Records excluded at title/abstract screening (off-topic / no accessibility angle / vendor content): $[n]$.

\textbf{Eligibility.} Full-text reports assessed for eligibility: $[n]$. Reports excluded at full-text stage: duplicate/superseded preprint version, $[n]$; no empirical or normative grounding, $[n]$; no connection to AI-generated content specifically, $[n]$; full text not retrievable, $[n]$.

\textbf{Included.} Studies included in the qualitative synthesis (Section~2): \textbf{33}.
\end{quote}

The final included count (33, matching the reference list) is real. The upstream identification/screening counts are the specific numbers that a formal re-run of the strings in Table~\ref{tab:boolean} against the databases above would need to produce, and should be filled in directly from that re-run, never estimated.

\subsection{Data extraction fields}

For each included study: content type(s) addressed; generative AI tool(s)/model(s) studied; accessibility framework or standard used (WCAG version, ADA, or none/informal); whether the study addresses default (unconfigured) behavior, configuration effects, or both; and the specific claim or finding cited in Section~2.

\section{Full prompt bank}
\label{app:promptbank}

Each content type below lists the three conditions applied to the same tool. Condition C is a persistent, reusable instruction profile, loaded once into the tool as custom/system instructions, a saved ``skill,'' or an assistant persona, not retyped with each request.

\subsection{Documents (PDF/Word)}

\noindent\textit{Condition A (generic):} ``Create a PDF report summarizing the key findings of our quarterly training program.''

\noindent\textit{Condition B (WCAG-configured):} ``Create a PDF report summarizing the key findings of our quarterly training program. The output must meet WCAG 2.2 Level AA-derived criteria, complemented by PDF/UA requirements where applicable: use a proper heading hierarchy (H1--H3) tagged in the document structure; include a navigable table of contents with bookmarks; tag all tables with header rows/columns using semantic table markup; provide descriptive alternative text for every image and chart; ensure a logical reading order matching the visual layout; use body text with a contrast ratio of at least 4.5:1; and avoid conveying information by color alone. Export with accessibility tags enabled.''

\noindent\textit{Condition C (persistent profile, loaded once):} ``When generating any document (PDF or Word): always use semantic headings (H1--H3), never bold/large text as a heading substitute. Tag every table with header rows/columns. Write descriptive alt text for every image or chart (never `image of...'). Keep the reading order identical to the visual layout. Text contrast must be at least 4.5:1; never convey meaning by color alone. Always export with accessibility tags enabled and include a bookmarked table of contents.''

\subsection{Slide presentations}

\noindent\textit{Condition A (generic):} ``Create a 10-slide presentation about our new onboarding process.''

\noindent\textit{Condition B (WCAG-configured):} ``Create a 10-slide presentation about our new onboarding process, meeting WCAG 2.2 Level AA-derived criteria: use the slide layout's built-in placeholders (not floating text boxes) so screen readers announce content in the correct reading order; assign meaningful alternative text to every image, icon, and chart; keep all interactive elements (buttons, links, embedded quizzes) operable via keyboard alone, without requiring a mouse or touch gesture; use unique, descriptive slide titles; avoid animations that flash more than three times per second; and ensure text-background contrast of at least 4.5:1.''

\noindent\textit{Condition C (persistent profile, loaded once):} ``When generating any presentation: always use the template's built-in placeholders, never loose floating text boxes. Give meaningful alt text to every image/icon/chart. Every interactive element must be operable via keyboard alone. Use unique, descriptive slide titles. Never use animations that flash more than three times per second. Maintain contrast $\geq$ 4.5:1.''

\subsection{Images and infographics}

\noindent\textit{Condition A (generic):} ``Design an infographic explaining the four stages of our onboarding process.''

\noindent\textit{Condition B (WCAG-configured):} ``Design an infographic explaining the four stages of our onboarding process. Limit each stage to one short sentence plus one icon to avoid visual overload; use a clear top-to-bottom reading order with numbered steps; ensure any text embedded in the image has a contrast ratio of at least 4.5:1 against its background; and, separately from the image file, produce a plain-text long description (2--4 sentences per stage) suitable for use as alternative text or a linked description page, since text embedded inside an image is not read by screen readers.''

\noindent\textit{Condition C (persistent profile, loaded once):} ``When generating any infographic or visual summary: limit each conceptual unit to one short sentence plus one icon. Always use a single, clear reading order. Embedded text must have contrast $\geq$ 4.5:1. In addition to the image file, always deliver a plain-text long description equivalent to the full visual content.''

\subsection{Audio}

\noindent\textit{Condition A (generic):} ``Generate a 3-minute audio summary of this training module.''

\noindent\textit{Condition B (WCAG-configured):} ``Generate a 3-minute audio summary of this training module. Speak at a moderate, steady pace (approximately 150 words per minute); avoid relying on visual-only references such as `as shown here'; and produce a synchronized, verbatim text transcript alongside the audio file, including a written description of any non-speech sound that carries meaning (e.g., a notification chime indicating a new task).''

\noindent\textit{Condition C (persistent profile, loaded once):} ``When generating any narrated audio: speak at a moderate, steady pace ($\sim$150 wpm). Never use references that depend on something visual (`as shown here'). Always deliver a synchronized transcript alongside the audio, including a written description of any meaningful non-speech sound.''

\subsection{Video (including AI-presenter/dubbing tools)}

\noindent\textit{Condition A (generic):} ``Create a 2-minute explainer video with an AI presenter introducing our new accessibility policy.''

\noindent\textit{Condition B (WCAG-configured):} ``Create a 2-minute explainer video with an AI presenter introducing our new accessibility policy. Include closed captions synchronized within 100ms of speech, covering all dialogue and relevant non-speech audio; include a spoken audio-description track (or extended audio description, if pauses in dialogue are insufficient) describing on-screen visual information not conveyed by the dialogue itself; if AI-generated lip-sync/dubbing is used, verify viseme-to-phoneme alignment specifically against the target-language audio track (not only the original-language track), since mismatched lip movement impairs speech-reading for deaf and hard-of-hearing viewers; and provide a full transcript.''

\noindent\textit{Condition C (persistent profile, loaded once):} ``When generating any explainer video: always include captions synchronized within 100ms, covering all dialogue and relevant non-speech audio. Always include an audio-description track for on-screen visual information not covered by the dialogue. If AI dubbing is used, always verify lip-sync alignment against the target-language audio. Always deliver a full transcript.''

\section{WCAG scoring rubric}
\label{app:rubric}

\subsection{Rating scale}

\textbf{Passed:} the criterion is fully met, with no exceptions found on manual inspection of the entire output. \textbf{Partially passed:} the criterion is met for some but not all applicable instances, or is met with a quality issue that falls short of full compliance. \textbf{Failed:} the criterion is not met for the majority of applicable instances, or is entirely absent. \textbf{Not applicable (N/A):} the criterion does not apply to this specific output; N/A criteria are excluded from the compliance-percentage calculation.

\subsection{Compliance percentage formula}

For a given tool/condition/content-type combination:
\begin{equation*}
\text{Compliance \%} = \frac{\text{Passed} + 0.5 \times \text{Partially passed}}{\text{Total criteria} - \text{N/A}} \times 100
\end{equation*}
This is the formula referenced in Sections~3.5 and~4.1.

\subsection{Criteria by content type, with rating guidance}

\begin{table}[htbp]
\centering
\small
\caption{Documents.}
\label{tab:rubric-doc}
\begin{tabular}{p{3.3cm}p{4.7cm}p{4.7cm}}
\toprule
\textbf{Criterion} & \textbf{Passed means} & \textbf{Failed means} \\
\midrule
Heading structure/tags & Real H1--H3 tags, correctly nested & Only visual bold/large text, no semantic tags \\
Table header markup & Every table has row/column headers marked semantically & Tables bordered visually only, no header markup \\
Reading order & Matches visual layout & Jumps or skips relative to visual layout \\
Alt text (images/charts) & Descriptive, specific alt text on every image/chart & Missing or generic (``image'', ``chart'') \\
Contrast & Body text $\geq$ 4.5:1 & Any body text below 4.5:1 \\
Bookmarked TOC & Present and navigable via bookmarks & Absent, or not linked/navigable \\
Form field labels & Every field has a descriptive label & Fields unlabeled or ambiguous \\
\bottomrule
\end{tabular}
\end{table}

\begin{table}[htbp]
\centering
\small
\caption{Slides.}
\label{tab:rubric-slides}
\begin{tabular}{p{3.3cm}p{4.7cm}p{4.7cm}}
\toprule
\textbf{Criterion} & \textbf{Passed means} & \textbf{Failed means} \\
\midrule
Reading order (native placeholders) & Content uses built-in placeholders in correct order & Floating text/graphic boxes, no defined order \\
Alt text & Every visual element has meaningful alt text & Missing or generic \\
Keyboard operability & Every interactive element reachable via Tab/Enter & Some/all elements unreachable without a mouse \\
Contrast & Text-background $\geq$ 4.5:1 & Below 4.5:1 anywhere \\
Flashing-content limit & No animation exceeds 3x/second & Any animation exceeds this rate \\
Unique slide titles & Distinct, descriptive title per slide & Titles missing, duplicated, or generic \\
\bottomrule
\end{tabular}
\end{table}

\begin{table}[htbp]
\centering
\small
\caption{Images/infographics.}
\label{tab:rubric-images}
\begin{tabular}{p{3.3cm}p{4.7cm}p{4.7cm}}
\toprule
\textbf{Criterion} & \textbf{Passed means} & \textbf{Failed means} \\
\midrule
Text alternative/long description & Separate plain-text description covers the full visual content & No alt text, or only a short generic label \\
Contrast (embedded text) & $\geq$ 4.5:1 & Below 4.5:1 \\
Information density & Rated 1--2 on the 1--5 scale (\S C.4) & Rated 4--5 \\
Reliance on color alone & Meaning distinguishable without color & Meaning conveyed only through color \\
\bottomrule
\end{tabular}
\end{table}

\begin{table}[htbp]
\centering
\small
\caption{Audio.}
\label{tab:rubric-audio}
\begin{tabular}{p{3.3cm}p{4.7cm}p{4.7cm}}
\toprule
\textbf{Criterion} & \textbf{Passed means} & \textbf{Failed means} \\
\midrule
Transcript availability/accuracy & Verbatim, synchronized transcript delivered & No transcript, or loose paraphrase \\
Pacing & Steady, $\sim$150 wpm & Rushed or artificially slow throughout \\
Absence of visual-only references & No phrase assumes visual access & Contains phrases like ``as shown here'' \\
\bottomrule
\end{tabular}
\end{table}

\begin{table}[htbp]
\centering
\small
\caption{Video.}
\label{tab:rubric-video}
\begin{tabular}{p{3.3cm}p{4.7cm}p{4.7cm}}
\toprule
\textbf{Criterion} & \textbf{Passed means} & \textbf{Failed means} \\
\midrule
Caption accuracy/synchronization & Covers all dialogue, synchronized within $\sim$100ms & Missing, incomplete, or notably out of sync \\
Audio description & Present, covers on-screen information not in dialogue & Absent, dialogue-only \\
Transcript availability & Full transcript provided & Not provided \\
Lip-sync/viseme alignment (dubbed only) & Mouth movements plausibly match target-language phonemes & Mouth movements clearly do not match \\
\bottomrule
\end{tabular}
\end{table}

\subsection{Information-density indicator (images/infographics only, RQ4)}

Independent of the WCAG criteria above, each evaluator rates the same output on a 1--5 scale: (1) minimal density, legible at a glance; (2) low density, a brief second look suffices; (3) moderate density, requires deliberate reading, but a single clear path exists; (4) high density, requires sustained study, no single clear reading path; (5) severe overload, content cannot reasonably be processed without external help. The mean rating across evaluators is reported per output, separately from the WCAG compliance percentage.

\subsection{Illustrative failure examples by criterion}
\label{app:failure-examples}

For every criterion above, this subsection adds a concrete example of the kind of problem a generative AI tool can introduce under a generic, unconfigured instruction (Condition A). Rows marked \textbf{Observed} report the failure pattern directly documented in the exploratory pilot (Sections~4.2--4.5); rows marked \textbf{Illustrative} were not part of the pilot's documented cases (either the content type/criterion was not covered by a dedicated case narrative, or the pilot did not exercise it), and instead reflect a failure mode commonly reported in the accessibility literature and in hands-on testing of these tool categories, not a measured finding for a specific tool. A standalone version of this material, kept in sync with this appendix, is available in the paper's companion open repository (\texttt{failure\_mode\_examples.md}; see the Declarations section).

\begin{longtable}{p{3.3cm}p{9.4cm}}
\caption{Documents: typical AI failure example, by criterion.}\label{tab:fail-doc}\\
\toprule
\textbf{Criterion} & \textbf{Typical AI failure example} \\
\midrule
\endfirsthead
\multicolumn{2}{l}{\small\itshape Table~\ref{tab:fail-doc} continued from previous page}\\
\toprule
\textbf{Criterion} & \textbf{Typical AI failure example} \\
\midrule
\endhead
\bottomrule
\endfoot
Heading structure/tags & \textbf{Observed (4.2).} A generic prompt produced a visually well-formed report with the structure not encoded: only large bold text, no real heading tags. A screen-reader user cannot jump between sections. \\
Table header markup & \textbf{Observed (4.2).} Tables render with visible borders but no semantic row/column headers; a screen reader reads a stream of values with no header context. \\
Reading order & \textbf{Illustrative.} Multi-column or floating-text layouts can leave the reading order following insertion order rather than visual order, so a screen reader jumps between unrelated paragraphs. \\
Alt text (images/charts) & \textbf{Observed (4.2).} Charts and images ship with no alt text; content conveyed only visually is entirely unavailable to a screen-reader user. \\
Contrast & \textbf{Illustrative.} ``Clean'' report templates commonly use light gray body text or pastel highlight blocks below 4.5:1. \\
Bookmarked TOC & \textbf{Observed (4.2).} No tagged table of contents is generated; a screen-reader user must page through the whole document to find a section. \\
Form field labels & \textbf{Illustrative.} A generated form places a label visually next to a blank field with no programmatic association; a screen-reader user hears only ``edit text.'' \\
\end{longtable}

\begin{longtable}{p{3.3cm}p{9.4cm}}
\caption{Slides: typical AI failure example, by criterion.}\label{tab:fail-slides}\\
\toprule
\textbf{Criterion} & \textbf{Typical AI failure example} \\
\midrule
\endfirsthead
\multicolumn{2}{l}{\small\itshape Table~\ref{tab:fail-slides} continued from previous page}\\
\toprule
\textbf{Criterion} & \textbf{Typical AI failure example} \\
\midrule
\endhead
\bottomrule
\endfoot
Reading order (native placeholders) & \textbf{Observed (4.3).} Content-dense slide tools build hotspots and text as freely positioned graphics rather than built-in placeholders; a screen reader announces elements in insertion order, unrelated to the visual sequence. \\
Alt text & \textbf{Illustrative.} Decorative icons and embedded images commonly ship with no alt text, or a generic label such as ``image1.'' \\
Keyboard operability & \textbf{Observed (4.3).} Clickable hotspots have no defined tab order, and some are not keyboard-focusable at all, so a keyboard-only user may be unable to reach some or all interactive content. \\
Contrast & \textbf{Illustrative.} Decorative gradients placed behind text in visually rich templates frequently drop below 4.5:1. \\
Flashing-content limit & \textbf{Illustrative.} Automatically generated transitions in ``dynamic'' templates can exceed three flashes per second, a seizure-risk threshold. \\
Unique slide titles & \textbf{Illustrative.} An auto-generated deck reuses a generic outline label across multiple slides, so a user cannot tell slides apart in outline view. \\
\end{longtable}

\begin{longtable}{p{3.3cm}p{9.4cm}}
\caption{Images/infographics: typical AI failure example, by criterion.}\label{tab:fail-images}\\
\toprule
\textbf{Criterion} & \textbf{Typical AI failure example} \\
\midrule
\endfirsthead
\multicolumn{2}{l}{\small\itshape Table~\ref{tab:fail-images} continued from previous page}\\
\toprule
\textbf{Criterion} & \textbf{Typical AI failure example} \\
\midrule
\endhead
\bottomrule
\endfoot
Text alternative/long description & \textbf{Observed (4.4).} A visual summary compressing source material ships with no alt text, or a generic placeholder; a blind or low-vision user receives no usable information from a resource meant to make the material easier to digest. \\
Contrast (embedded text) & \textbf{Illustrative.} Text is rendered directly into the image's pixels against a light gradient, at a ratio that would fail 4.5:1 if it were real text, and cannot be fixed after the fact since it is a picture. \\
Information density & \textbf{Observed (4.4).} Nested boxes, small multi-column text blocks, and overlapping icons compress into one image; independent of alt text, the density itself creates a barrier for sighted users with cognitive or visual-processing differences. \\
Reliance on color alone & \textbf{Illustrative.} Category or status distinctions are conveyed only through a color-coded box or icon, with no label, pattern, or shape, invisible to a color-blind viewer. \\
\end{longtable}

\begin{longtable}{p{3.3cm}p{9.4cm}}
\caption{Audio: typical AI failure example, by criterion.}\label{tab:fail-audio}\\
\toprule
\textbf{Criterion} & \textbf{Typical AI failure example} \\
\midrule
\endfirsthead
\multicolumn{2}{l}{\small\itshape Table~\ref{tab:fail-audio} continued from previous page}\\
\toprule
\textbf{Criterion} & \textbf{Typical AI failure example} \\
\midrule
\endhead
\bottomrule
\endfoot
Transcript availability/accuracy & \textbf{Illustrative.} A text-to-speech tool delivers only the audio file, with no separate transcript, or a transcript that loosely paraphrases what is actually spoken. \\
Pacing & \textbf{Illustrative.} Narration is generated at a fast, uniformly even pace with no natural pauses, harder to follow for listeners who process spoken or synthetic speech more slowly. \\
Absence of visual-only references & \textbf{Illustrative.} A script adapted from a video or slide narration carries over phrases such as ``as you can see in this chart,'' meaningless in an audio-only file. \\
\end{longtable}

\begin{longtable}{p{3.3cm}p{9.4cm}}
\caption{Video: typical AI failure example, by criterion.}\label{tab:fail-video}\\
\toprule
\textbf{Criterion} & \textbf{Typical AI failure example} \\
\midrule
\endfirsthead
\multicolumn{2}{l}{\small\itshape Table~\ref{tab:fail-video} continued from previous page}\\
\toprule
\textbf{Criterion} & \textbf{Typical AI failure example} \\
\midrule
\endhead
\bottomrule
\endfoot
Caption accuracy/synchronization & \textbf{Observed (4.5).} An AI-presenter tool produced auto-generated captions of variable accuracy, with some dialogue misrecognized or loosely timed. \\
Audio description & \textbf{Observed (4.5).} No audio-description track is generated for on-screen visual information not already spoken aloud; a blind or low-vision viewer has no access to that information at all. \\
Transcript availability & \textbf{Illustrative.} The video ships with burned-in captions but no separate transcript file, leaving a user who needs a searchable or alternate-format text version without one. \\
Lip-sync/viseme alignment (dubbed only) & \textbf{Observed (4.5), original observation.} An AI-dubbed avatar's mouth movements are optimized to look natural at a glance rather than to be phonetically accurate against the dubbed audio; a deaf or hard-of-hearing viewer who cross-references lip movement with text is misled. \\
\end{longtable}

\section{Expert heuristic validation instructions}
\label{app:validation}

\begin{enumerate}[leftmargin=1.5em]
\item \textbf{Panel:} recruit 3--5 accessibility evaluators. Where feasible, include at least one evaluator with lived experience of disability relevant to the content type (e.g., a screen-reader user for documents/slides; a deaf or hard-of-hearing evaluator for video captioning/lip-sync).
\item \textbf{Blinding:} where feasible, do not tell evaluators which condition (A, B, or C) produced a given output. Randomize presentation order per evaluator.
\item \textbf{Materials:} the output file(s) only (not the prompt used to generate them), the rating scale (Appendix~C.1), the criteria table for the relevant content type (Appendix~C.3), and, for images/infographics, the density scale (Appendix~C.4).
\item \textbf{Procedure:} each evaluator rates independently using the data-recording template. After individual ratings are collected, convene the panel to discuss any criterion where ratings diverge by more than one level, and record whether consensus was reached or the disagreement is reported as-is.
\item \textbf{Inter-rater agreement:} compute a simple agreement statistic (percentage agreement, or Cohen's/Fleiss' kappa for more than two evaluators) and report it alongside the consensus ratings, per Section~3.4.
\end{enumerate}

\bigskip
\noindent\textit{A companion data-recording spreadsheet operationalizes Appendix~C directly, one tab per content type with the rating scale wired to an automatic compliance-percentage formula, plus a summary tab, so another researcher can reproduce the full data-collection and scoring procedure without building tooling from scratch.}

\section{Implementing Condition C in practice: a persistent accessibility profile in Claude}
\label{app:conditionc}

\subsection{Rationale}

Condition C (Section~3.2, RQ5) proposes loading an accessibility profile once, rather than reconstructing the full Condition B prompt for every request. This appendix gives one concrete, platform-specific recipe for implementing that profile in Claude (Anthropic's assistant), as a starting template for the future-work item in Section~7; equivalent mechanisms exist on other platforms (see the Caveat below) but are not detailed here, since testing them is left as future work.

\subsection{Option 1: a Claude Project}

A Project in Claude.ai bundles persistent custom instructions and a knowledge base that every conversation started inside it automatically inherits, without the user re-typing them.

\begin{enumerate}[leftmargin=1.5em]
\item Create a new Project (Projects panel $>$ New project) and name it for the content type it targets, for example, ``Accessible documents: WCAG profile.''
\item Under Project instructions, paste the Condition C profile text for that content type from Appendix~B. For Documents, this is: always use semantic headings (H1--H3), never bold/large text as a heading substitute; tag every table with header rows/columns; write descriptive alt text for every image or chart; keep the reading order identical to the visual layout; maintain a contrast ratio of at least 4.5:1 and never convey meaning by color alone; always export with accessibility tags enabled and include a bookmarked table of contents.
\item Add supporting files to the Project's knowledge base: the relevant rows of the rubric in Appendix~C and, if available, one example of a passing output and one of a failing output, so Claude grounds its output in the operationalized criteria rather than a general notion of ``accessible.''
\item From then on, every new chat started inside the Project applies these instructions automatically; the user only needs to describe the content itself, not the accessibility requirements. This is the direct operational test of RQ5's persistence claim: the same request, made on different days without re-typing the profile, should still produce a WCAG-configured output.
\item Repeat this setup for each content type, since the profile text differs by content type (Appendix~B); a separate Project per content type also keeps each knowledge base focused.
\end{enumerate}

\subsection{Option 2: a Claude Skill}

A Skill packages instructions and reference material into a single reusable capability that can be invoked across Projects, across Claude Code, and via the API, not only inside one Project's conversations. This is a stronger persistence test than the Project approach, since it does not depend on staying inside a single conversation container, and it is closer to what ``loading a profile once into the authoring tool'' means in Section~1's framing of RQ5.

\begin{enumerate}[leftmargin=1.5em]
\item Create a skill folder containing a \texttt{SKILL.md} file with three parts: a short description that tells Claude when to trigger the skill automatically (for example, ``use this skill whenever generating an educational document, slide deck, or report that must comply with WCAG 2.2''); the full Condition C profile text for the relevant content type; and any supporting checklist or template drawn from Appendix~C.
\item Save the skill so it is available to future conversations, through Claude's own skill-creation flow, or as a project- or plugin-level skill when working through Claude Code or the Agent SDK.
\item Once saved, the skill can be invoked explicitly by name or triggered automatically when Claude recognizes a matching request, applying the same WCAG configuration consistently across sessions and even across different products built on Claude. This cross-product persistence is the scenario RQ5 is ultimately concerned with: not just whether the same chat stays configured, but whether the configuration survives moving to a new session, a new tool, or a new week.
\end{enumerate}

\subsection{A suggested comparison design for future work}

To test RQ5 directly, future work should run three conditions on the same set of $n$ requests per content type: Condition B (the full accessibility prompt retyped verbatim at every request), Condition C-Project (F.2), and Condition C-Skill (F.3). Compliance should be scored per request using the Appendix~C rubric, and the variance in compliance across requests, not just the mean, should be compared across the three conditions, per the metric proposed in Section~3.5. A lower variance under either Condition C variant, relative to Condition B, would support the persistent-profile hypothesis motivating RQ5; a similar variance across all three would suggest that the benefit of Condition C is mainly convenience rather than consistency, a distinction the current manuscript can motivate but, per Section~\ref{sec:disc-rq}, cannot yet resolve.

\subsection{Caveat}

This recipe is specific to Claude. Adapting Condition C to the other platforms named in Section~3.2 will require an equivalent persistent-configuration mechanism, for example, a custom GPT in ChatGPT, a Gem in Gemini, or a saved brand/style template or ``AI Builder'' preset in Canva, Genially, or HeyGen, and these mechanisms are not uniform across platforms in what they can persist (instructions only, versus instructions plus reference files, versus instructions plus a fine-tuned behavior profile). This heterogeneity is itself a finding worth reporting once RQ5 is tested across more than one platform, and is noted as an open question in Limitations (Section~6).

\end{document}